# The Identification of Thresholds and Time Delay in Self-Exciting Threshold AR Model by Wavelet

Song-Yon Kim and Mun-Chol Kim

Faculty of Mathematics, **Kim Il Sung** University, Pyongyang, DPRK

**Abstract**: In this paper we studied about the wavelet identification of the thresholds and time delay for more general case without the constraint that the time delay is smaller than the order of the model. Here we composed an empirical wavelet from the SETAR (Self-Exciting Threshold Autoregressive) model and identified the thresholds and time delay in the model using it.

**Key words**: threshold autoregressive model, threshold, time delay, wavelet

## 1. SETAR and Problem

Generally, the thresholds and time delay in the time series model are estimated using LSM (Least Square method) and AIC [1, 2, 3, 4, 8].

And recently some identification of thresholds and time delay by wavelet is suggested under the assumption that the time delay is smaller than the order of the model for TAR and DTARCH [5, 6, 7, 9].

SETAR model was introduced in 1978 and applied to many fields such as modeling of Sun spots data [4, 6, 7, 8, 10].

The model on probability space ($\Omega$, $F$, $P$) is as follows.

$$x_t = \sum_{l=1}^{r+1}\left( b_{l0} + \sum_{m=1}^{p_l} b_{lm} + \varepsilon_{lt} \right)\mathbf{1}(x_{t-d} \in (\lambda_{l-1}, \lambda_l]). \tag{1}$$

Here $\{\varepsilon_{lt}, t = 0, 1, 2, \dots\}$ are i.i.d random variables with mean zero and varience $\sigma_l^2$ for each $l = \overline{1,\ t+1}$ and $\{\varepsilon_{lt}, t = 0, 1, 2, \dots\}, l = \overline{1,\ t+1}$ are mutually independent, $\lambda_0 = -\infty$, $\lambda_{r+1} = +\infty$.

The model (1) is called **SETAR** ($\boldsymbol{d}$, $\boldsymbol{r}$, $\boldsymbol{p_1}$, $\boldsymbol{p_2}$, …, $\boldsymbol{p_{r+1}}$) with time delay $\boldsymbol{d}$ , thresholds $\boldsymbol{r}$ and orders $\boldsymbol{p_1}$, $\boldsymbol{p_2}$, …, $\boldsymbol{p_{r+1}}$ .

Let $p = \max\{\boldsymbol{p_1}, \boldsymbol{p_2}, \dots, \boldsymbol{p_{r+1}}\}$ where $\boldsymbol{p_1}, \boldsymbol{p_2}, \dots, \boldsymbol{p_{r+1}}$ are known.

And let $b_{lm} = 0$ when $p_l < m \le p$ then (1) can be written as

$$x_t = \sum_{l=1}^{r+1}\left( b_{l0} + \sum_{m=1}^{p} b_{lm} x_{t-m} + \varepsilon_{lt} \right) \mathbf{1}_{(x_{t-d} \in (\lambda_{l-1},\ \lambda_l])} \tag{2}$$

where $d$ is an bounded integer(constant) which is not related to $p$, $d \le D < \infty$.

Then we consider the wavelet identification of $d$ and $r, \lambda_1, \cdots, \lambda_r$ using a specific empirical wavelet.

## 2. Empirical Wavelet

**Assumption 1** Wavelet $\psi(x),\ x \in \mathbf{R}$ satisfies the following conditions.

i) Compactly supported on $[-A,\ A]$ with $A > 1$.

ii) $\psi$ is of bounded variation on $[-A,\ A]$.

iii) $\psi(x) = 0,\ x \in [-1,\ 1]$

$$\int_{-A}^{A} \psi(x)dx = 0,\quad \int_{-A}^{A} x\psi(x)dx = 0,\quad \int_{-A}^{A} \psi^2(x)dx < \infty,\quad \int_{1}^{A} \psi(x)dx \ne 0,\quad \int_{1}^{A} x\psi(x)dx \ne 0$$

From the wavelet $\psi(x)$ satisfying Assumption1 and a scale function $\phi(x)$, we can obtain an orthogonal wablet basis on $L_2[a,\ b]$ as follows.[10].

$$\{\phi_{l,k}^{per},\ k \in I_l,\ \psi_{j,k}^{per},\ k \in I_j,\ j \ge l\},\quad I_j = \{0,\ 1,\ \cdots,\ 2^{j-1}\},\ j = 1,\ 2,\ \cdots$$

where

$$\phi_{l,k}^{per}(x) = \sum_{n=-\infty}^{+\infty} (b-a)^{-1/2} \phi_{l,k}\left( \frac{x-a}{b-a} + n \right),\quad \psi_{j,k}^{per}(x) = \sum_{n=-\infty}^{+\infty} (b-a)^{-1/2} \psi_{j,k}\left( \frac{x-a}{b-a} + n \right),$$

$$\phi_{l,k}(x) = 2^{l/2}\phi(2^l x - k),\quad \psi_{j,k}(x) = 2^{l/2}\psi(2^j x - k).$$

Then for $f \in L_2[a,\ b]$,

$$f(x) = \sum_{k \in I_l} \alpha_{l,k} \phi_{l,k}^{per}(x) + \sum_{j \ge l} \sum_{k \in I_j} \beta_{j,k} \psi_{j,k}^{per}(x) \tag{3}$$

where $\alpha_{l,k} = \int_a^b \phi_{l,k}^{per}(x) f(x)dx$, $\beta_{j,k} = \int_a^b \psi_{j,k}^{per}(x) f(x)dx$.

The expression (3) is called the wavelet expansion of $f$ on $[a,\ b]$ and $\beta_{j,k}$ is called the wavelet coefficient of $f$. And Model (2) is considered under the following assumption.

**Assumption 2** i) $x_t$ is geometrically ergogic.

ii) The probability density function $g(x_1, x_2, \cdots, x_p, \cdots, x_D)$ and $f$ of $\varepsilon_{lt}$ are bounded on $(-\infty,\ +\infty)$ and $\mathbf{R}^D$ respectively and there exists $\mathrm{U}$, $[a,\ b] \subset \mathrm{U}$ such that

$$M_1 \le g(x_1,\ \cdots,\ x_D),\ f(x) \le M_2,\ x \in \mathrm{U},\ (x_1,\ \cdots,\ x_D) \in \mathrm{U}^D$$

where $M_1,\ M_2 > 0$ are constants.

iii) $\varepsilon_{lt}$ are only on $[-M,\ M]$. $0 < M < \infty$.

Let $H(x_1, x_2, \cdots x_p\,;\, x_d) := \sum_{l=1}^{r+1}\left( b_{l0} + \sum_{m=1}^{p} b_{lm} x_m \right) \mathbf{I}_{(\lambda_{l-1},\ \lambda_l]}(x_d)$.

Then model (2) has the following form:

$$x_t = H(x_{t-1}, x_{t-2}, \cdots, x_{t-p}\,; x_d) + \sum_{l=1}^{r+1} \mathbf{I}_{(\lambda_{l-1},\, \lambda_l]}(x_d)\varepsilon_{lt}$$

where $\{x_t\}_{t=0,1,2,\cdots}$ are sampled from SETAR model (2).

Let $X_{l-1}^i := \begin{cases} (x_{l-1}, x_{l-2}, \cdots, x_{l-p}, a, \cdots, a), & 1 \le i \le p \\ (x_{l-1}, x_{l-2}, \cdots, x_{l-p}, a, \cdots, a, x_{l-i}, a, \cdots, a), & p < i \le D \end{cases}.$

For any $T = (t_1, \cdots, t_D)$, $\omega \in \Omega$ let

$$N_n^i(T,\ \omega) := \{l : D+1 \le l \le n,\ \| X_l^i(\omega) - T \| < \delta_n,\ X^i(\omega) \le T\}$$

where $\|\cdot\|$ is the Euclidean norm.

Let $\delta_n := n^{-1/(D+2)}$, $\mathrm{I}(s,\ \delta) := \{k : | a + k(b-a)/2^j - s \le \delta |\}$,

$$\mathrm{R}_{j^*} := \{(t_{j^*1}, t_{j^*2}, \cdots, t_{j^*(D-1)}) : t_{j^*l} = a + k_l(b-a)/2^{j^*},\ k_l \in \mathrm{I}_{j^*}\},\ j^* = 1, 2, \cdots,$$

$$T_{j^*}^{m,\,s} := (t_{j^*1}, t_{j^*2}, \cdots, t_{j^*(m-1)}, s, t_{j^*m}, \cdots, t_{j^*(D-1)}),\ \ \forall T_{j^*} \in \mathrm{R}_{j^*},$$

$$s_i := a + i(b-a)/N,\ i = 1, 2, \cdots\ ,\ n_{m_i}(\omega) := \# N_N^m(T_{j^*}^{m,\,s_i},\ \omega),\ N := [n^{1/(D+2)}].$$

For any $T_{j^*} \in \mathbf{R}_{j^*}$, the empirical wavelet

$$W_{j,k}^m(T_{j^*},\ \omega) = \frac{b-a}{N}\sum_{i=1}^{N} \psi_{j,k}^{per}(s_i)\frac{1}{n_{m_i}(\omega)}\sum_{l \in N_n^m(T_{j^*}^{m,\,s_i},\ \omega)} x_l(\omega),\ \ m = 1, 2, \cdots, p,\ p+1, \cdots, D,\ \ j = 1, 2, \cdots,\ k \in I_j$$

is composed.

### 3. Main Result

**Theorem** Under the assumption 1, 2. $\lim_{n\to\infty} 2^{3j_n}/N = 0$ as $n \to \infty$, $j_n \to \infty$ for $j_n$.

Then there exists a enough large $j^* > 0$ such that the followings hold.

① When $K \in \bigcup_{l=1}^{r} I(\lambda_l,\ 2^{-j_n}(b-a))$ there exists $\mathrm{T}_{j^*}^0$, such that

$$\lim_{n\to\infty} P\{| W_{j,\,k}^d(\mathrm{T}_{j^*}^0) | \ge c_0 2^{-3j_n/2}\} = 1.$$

② When $K \notin \bigcup_{l=1}^{r} I(\lambda_l,\ 2^{-j_n}(b-a))$ there exists $\mathrm{T}_{j^*}^0$, such that

$$\lim_{n\to\infty} P\{| W_{j_n,\,k}^d(\mathrm{T}_{j^*}^0) | = \circ(2^{-2j_n})\} = 1,\ \ W_{j_n,\,k}^d(\mathrm{T}_{j^*}^0) = \circ(2^{-2j_n})\ \text{(a.s).}$$

③ When $m \ne d$, $\max_{k \in I_{j_n},\ \mathrm{T}_{j^*} \in \mathbf{R}_{j^*}} | W_{j_n,\,k}^m(\mathrm{T}_{j^*}) | = \circ(2^{-2j_n})$ (a.s).

From the theorem we can calculate the empirical wavelet $w_{j,\,k}^m$ for different $m$ at different levels of $j$ and take $m$ that gives $w_{j,\,k}^m$ large absolute value –the "peak" in its range as time delay. Next, we calculate $w_{j,\,k}^m$ at the identified $m$ for different $k$ and take $k$ that gives $w_{j,\,k}^m$ high peaks around them as thresholds.